\documentclass[manuscript]{acmart}

\AtBeginDocument{%
  }

\usepackage{multirow}
\usepackage{diagbox}

\copyrightyear{2023}
\acmYear{2023}
\setcopyright{rightsretained}
\acmConference[RecSys '23]{Seventeenth ACM Conference on Recommender Systems}{September 18--22, 2023}{Singapore, Singapore}
\acmBooktitle{Seventeenth ACM Conference on Recommender Systems (RecSys '23), September 18--22, 2023, Singapore, Singapore}
\acmDOI{10.1145/3604915.3610660}
\acmISBN{979-8-4007-0241-9/23/09}

\begin{document}

\title{A Model-Agnostic Framework for Recommendation via Interest-aware Item Embeddings}


\author{Amit Kumar Jaiswal}
\orcid{0000-0001-8848-7041}
\affiliation{%
  \institution{University of Surrey}
  \country{United Kingdom}
}
\email{a.jaiswal@surrey.ac.uk}

\author{Yu Xiong}
\affiliation{%
  \institution{University of Surrey}
  \country{United Kingdom}}
\email{y.xiong@surrey.ac.uk}

\renewcommand{\shortauthors}{Amit and Yu}

\begin{abstract}
Item representation holds significant importance in recommendation systems, which encompasses domains such as news, retail, and videos. Retrieval and ranking models utilise item representation to capture the user-item relationship based on user behaviours. While existing representation learning methods primarily focus on optimising item-based mechanisms, such as attention and sequential modelling. However, these methods lack a modelling mechanism to directly reflect user interests within the learned item representations. Consequently, these methods may be less effective in capturing user interests indirectly. To address this challenge, we propose a novel Interest-aware Capsule network (IaCN) recommendation model, a model-agnostic framework that directly learns interest-oriented item representations. IaCN serves as an auxiliary task, enabling the joint learning of both item-based and interest-based representations. This framework adopts existing recommendation models without requiring substantial redesign. We evaluate the proposed approach on benchmark datasets, exploring various scenarios involving different deep neural networks, behaviour sequence lengths, and joint learning ratios of interest-oriented item representations. Experimental results demonstrate significant performance enhancements across diverse recommendation models, validating the effectiveness of our approach.
\end{abstract}




\maketitle

\section{Introduction and Prior Work}\label{sec:intro}
The field of recommendation systems has witnessed significant advancements in recent years, primarily due to the rapid progress of deep learning techniques. Notable successes have been observed in various domains, including news recommendation, e-commerce, video recommendation, and advertisement. In the context of recommendation systems, it is essential to consider user interaction behaviours as they convey valuable information about the user's interests. These behaviours indicate not only the individual items within the sequences but also the user's singular or multiple interests. Users often exhibit diverse interests, engaging with products from distinct categories such as sports, food, and clothing. Identifying these interests becomes challenging as they are embedded within the interactive behaviours, rendering direct capture of such (intrinsic) interests set forth a formidable task. The primary objective in recommendation systems revolves around effectively targeting the diverse interests of varied users. Existing deep learning models focussed on click-through rate (CTR) prediction such as Wide and Deep models~\cite{cheng2016wide} combine the training of wide linear models and deep neural networks, thereby harnessing the advantages of both memorisation and generalisation for recommender systems as opposed to self-attention based models~\cite{cheng2021looking}. Deep Crossing~\cite{shan2016deep} with follow up deep and cross models~\cite{wang2017deep,wang2021dcn}, Product-network in network (PIN)~\cite{qu2018product}, DeepFM~\cite{guo2017deepfm}, adaptive factorization network~\cite{cheng2020adaptive} and xDeepFM~\cite{lian2018xdeepfm} aim to extract low-order and explicit high-order features by incorporating a product layer, and traditional factorization machine module. In the case of deep interest network (DIN)~\cite{zhou2018deep}, an attention mechanism is employed to enhance the pooling weights of similar items. However, the deep interest evolution network (DIEN)~\cite{zhou2019deep} introduces a sequential model to capture the sequential characteristics instead of relying solely on item embeddings as in DIN. DIEN utilises the hidden states of the GRU as inputs for attention and employs attentional update GRU (AUGRU) as a replacement for the conventional attention model. Overall, deep neural networks play a crucial role in depicting user interests based on past user-item interactions~\cite{pal2020pinnersage,liu2020octopus}, utilising item embedding vectors. To address the issue pertaining to the dispersion of user interests, a method known as the dynamic routing capsule network~\cite{sabour2017dynamic} proposes to mitigate the limitations of traditional convolutional neural networks (CNNs), which successfully achieves a more comprehensive understanding of the relationships between objects~\cite{edraki2020subspace}. Building upon the concept of dynamic routing,~\cite{li2019multi} propose the multi-interest network with dynamic routing to handle the diverse interests of users in retrieval systems. This method adopts a strategy of representing a single user using multiple vectors, each encoding distinct aspects of the user's diverse interests. By employing this multi-vector representation approach,~\cite{li2019multi} enhances the capacity to capture and account for the various facets of user interests in the retrieval process. Nonetheless, current representation learning methods primarily concentrate on optimising the item-based mechanism that connects user interactive behaviour sequences with candidate items. Consequently, they overlook the crucial factor of label diffusion within each item present in the user interactive behaviour sequences. This oversight weakens the impact of user interests during the backpropagation phase of the training model. Additionally, as network layers become deeper and the dimension of the embedding layer increases, the introduction of new methods often necessitates the redesign of the entire model architecture or the inclusion of additional datasets or information. Hence, it becomes imperative to develop a method that can enhance model performance while maintaining generalisability, eliminating the need for extensive architectural modifications or the introduction of supplementary information. In this work, a novel framework, Interest-aware Capsule network (IaCN), is proposed that can effectively learn item representations based on user interests. Our proposed framework directly incorporates the concept of user interests to enhance the modelling process. To ensure the generalisability of the framework, an auxiliary task to cater users' multiple interests based on~\cite{pi2019practice} has been proven effective in improving model performance. By jointly learning item-based and interest-based item representations, the framework enables the integration of diverse user interests into the entire model. Consequently, the interest-based item representations generated by IaCN are shared with the original model, thereby enriching the overall information captured by the model.\\
\textbf{Our Contributions:} To summarise, we make the following contributions:
\vspace{-4pt}
\begin{itemize}
    \item We propose a novel framework, the User Interest-aware Capsule network, which aims to facilitate the learning of interest-based item representations. By incorporating IaCN as an auxiliary task, this framework can be seamlessly integrated with existing recommendation models.
    \item We propose an approach that combines a joint learning method and hyperparameter optimisation within the IaCN framework. Through extensive experimentation on benchmark datasets, our results demonstrate significant improvements across various existing recommendation models.
\end{itemize}

\section{The Proposed Model}
Within the context of recommendation systems, $i$ represents the identification of an item, and the set of user interactive behaviour sequences, represented by $\mathbb{S}_u$, encompasses the sequences of items that a user $u$ has either clicked on or viewed. The symbol $u_p$ corresponds to the user's basic profile information. The identification of a candidate item $i_t$, while $r_t$ represents the associated information of said candidate item, obtained from the recommendation system. The function $f(i_t,r_t)$ represents the feature function that captures the characteristics of the candidate item information derived from the combination of the candidate item $i_t$ and the item-related information $r_t$. The representation of user interests is typically performed to facilitate the learning of the function $f(\mathbb{S}_u, u_p)$, which encompasses both the user profile and the user's interactive behaviour sequences. Therefore, the formulation of user interests can be expressed as follows,
\vspace{-8pt}
\begin{align}
    \mathcal{I}_u = f(\mathbb{S}_u, u_p)\mid\mathcal{I} = (\mathrm{i}^{1}_u, \mathrm{i}^{2}_u,\ldots,\mathrm{i}^{K}_u)\in\mathbb{R}^{h\times K}
\end{align}
where $\mathcal{I}$ denote the representation vector that is learned from the user information $\mathbb{S}_u$ and $u_p$. The dimension of the user interest vector is denoted by $h$, while $K$ represents the number of dimensions in the user vector. Specifically, the vector $\mathrm{i}^{1}_u$ represents one of the multiple interest vectors of the user, and $\mathcal{I}$ represents the collection of these interest vectors.
\begin{figure*}[h!]
\centering
  \includegraphics[width=\linewidth,height=6cm]{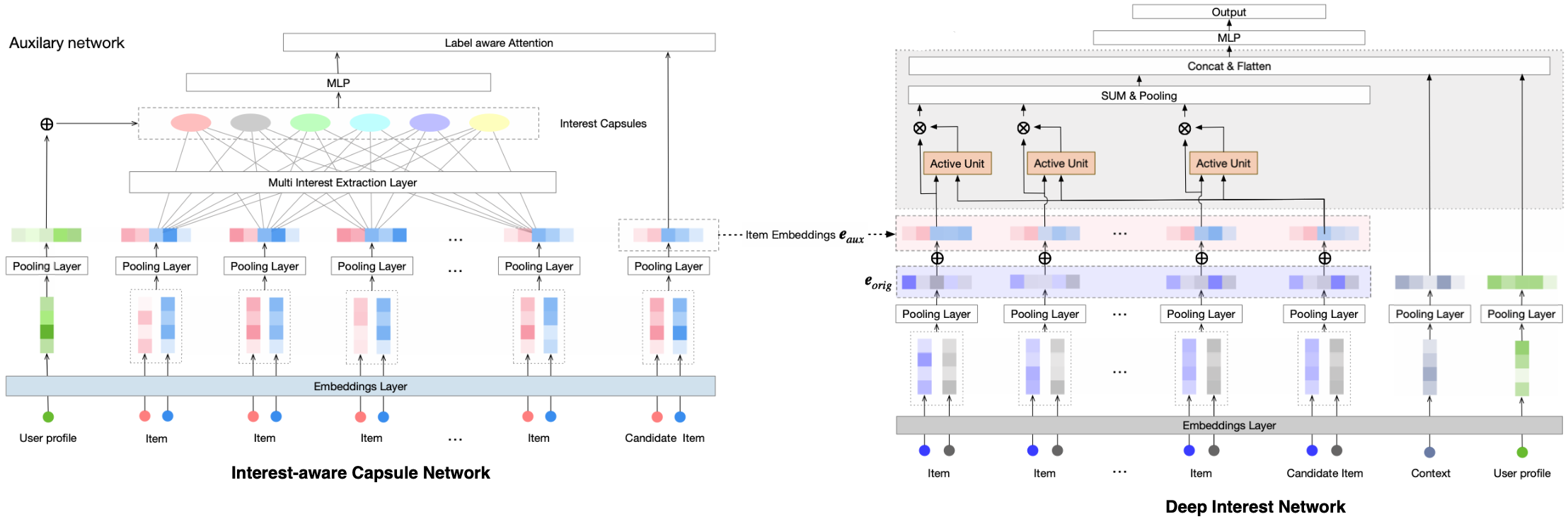}
  \caption{A schematic architecture of the IaCN framework which serves as an auxiliary task and shares item embeddings with the deep interest network. Within the IaCN, user interactive behaviour sequences and user profiles are incorporated and processed through a dynamic routing capsule mechanism. This mechanism enables the extraction of user interests in the form of capsules. We employ a scaled dot product between the interest capsules and candidate item embeddings, these elements are brought into alignment within the same representation space.}\label{fig:recsys}
\end{figure*}
In addition, the representation of candidate items aims to learn the function $f(i_t,r_t)$, which involves the item id and its associated information. The candidate item embeddings can be obtained through $c_t = f(i_t,r_t)\mid c_{t}\in\mathbb{R}^h$, where vector $c_t$ represents the embedding representation learned from the target item id $i_t$ and its corresponding associated information $r_t$. Specifically, the vector $c_t$ consistently refers to the vector obtained from the embeddings and pooling layer in the Interest-aware Capsule Networks, as discussed in the subsequent section. The purpose of a recommendation system is to promote `desirable' items to users navigating a website or web application. Consequently, it is essential to establish a scoring mechanism that measures the relationship between a candidate item and the user's interests. This score can be defined as $\mathcal{R}_{\text{score}}(\mathcal{I}_u, c_t) = c^{T}_t \mathrm{i}^{K}_u$. By calculating the value of $\mathcal{R}_{\text{score}}$, we can quantitatively assess the proximity between user interests and candidate items. Ultimately, the recommendation system will utilise the collection of users multiple interests to identify and present the most desirable items to the user.
\vspace{-6pt}
\subsection{Interest-aware Capsule Network}
To overcome certain limitations associated with convolutional neural networks, capsules have been introduced as an alternative that employs vector-based representations instead of scalars to encode appearance features~\cite{sabour2017dynamic}. This transition enables improved preservation of spatial relationships among complete objects and their constituent parts. Additionally, capsules incorporate the dynamic routing mechanism, which enables the weighing of part contributions to the entire object in a varying manner at each inference step. Users' diverse interests are typically concealed within their interactive behaviour sequences and profile information. Capsules~\cite{sabour2017dynamic} offer a means to encode appearance feature representations by employing vectors instead of scalars, accomplished through the assembly of a group of neurons. Through dynamic routing in capsule networks, the weights of different capsules are learned, enabling the encoding of relationships between parts and wholes. Notably, capsules demonstrate a superior comprehension of object relationships compared to CNNs~\cite{edraki2020subspace}. In the domain of recommendation systems, multi-interest with dynamic routing network~\cite{li2019multi} leverages dynamic routing of capsules to automatically capture high-level multiple user interests, resulting in commendable performance within the e-commerce retrieval system. Building upon this foundation, we propose an item representation approach utilising user interest-aware capsule networks (IaCN) to enhance the performance of CTR prediction models. We follow the notion of the dynamic routing mechanism for Behaviour-to-Interest routing~\cite{li2019multi}, employed by capsules for the purpose of learning multi-interest representations based on user profile information and interactive behaviour sequences. The input to each capsule, $\mathrm{i}_j = \|s_j \|^2 / (1+ \|s_j \|^2) (s_j / \|s_j \|)$, here, $\mathrm{i}_j$ represents the output, while $s_j$ encompasses all the inputs of capsule $j$. The formulation of components are,
\begin{align}
    s_j = \sum_{i,j}a_{ij}\hat{x}_{j\mid i} = \sum a_{ij} W_{ij} x_{j}, a_{ij}=\frac{\exp(b_{ij})}{\sum_{k}\exp(b_{ik})}
\end{align}
Subsequently, by employing dynamic routing, the aim is to extract high-level abstract interests from the raw user features. The function $a_{ij}$ represents the activation operation (Softmax) applied to the input $b_{ij}$. The Behaviour to Interest approach selectively combines the user's viewing sequences into vectors representing multiple interests in an adaptive manner. Based on the routing logits ($\mathcal{R}_{\text{score}}$), the variable $b_{ij}$ can be expressed as, $b_{ij} = u^{T}_j M c_i \mid i\in I^c \forall \{0,1,\ldots,n\}, j\in \{0,1,\ldots,K\}$, where $c_i \in \mathbb{R}^h$ represents an embedding vector of item $i$ derived from the user's interactive behaviour sequences. The variable $v{c_i}$ denotes one specific item embedding vector from these sequences. Moreover, $u_j \in \mathbb{R}^h$ refers to the capsule vector representing the user's interests, where $j$ ranges from 1 to $K$. The hyperparameter $K$ represents the number of user interests. The matrix $M\in\mathbb{R}^{h\times h}$, facilitates the bilinear mapping and serves as a linkage between the user's capsule interests and the viewed sequences. Meanwhile, $b_{ij}$ represents the connection between the user's interest and the item, ensuring they are mapped to the same vector space. While multi-interest dynamic routing network~\cite{li2019multi} captures multi-interest capsule vectors from user interactive behaviour sequences and profile information, it also introduces label-aware attention using the scaled dot product to assess the relationship between user interests and item information. Within the label-aware attention layer (as shown in Figure~\ref{fig:recsys}), the candidate item serves as the query, the user interest capsule acts as the key and value, and the candidate item embedding vector is represented within the interest capsule space. The label-aware attention layer incorporates the item embedding vector along with the item embeddings from the original recommendation model to incorporate user-diverse, interest-based item representations into the primary model. The primary model is specifically designed to accommodate the requirements of the particular task at hand, such as a ranking model (DIN and DIEN). The formulation of the scaled dot product can be expressed as $\mathrm{i}_u = \mathcal{I}_u$ \text{Softmax}(\text{pow}($\mathcal{I}^{T}_u c_i, p))$, where $p$ refers to the user profile information. As a result, the probability $P(c_i \mid\mathrm{i}_u)$ is derived, and the softmax activation function is applied to select the most desirable option. The training loss is formulated as $\mathcal{L}_{\text{IaCN}} = \sum_{u,i}\log P(c_{i}\mid\mathrm{i}_u)$, where $\mathcal{L}_{\text{IaCN}}$ represents the loss incurred by the user interest-aware capsule network. It is crucial to ensure compatibility between the user's interests and the candidate item. To achieve this, the item embedding vector and the user's interest capsule vector are required to share a common vector space, which is established based on the representation of the user's interests. This alignment is of significant importance when aiming to construct interest-driven item representations.

\subsection{Augmentation of Item Embeddings with User Interests}
The embedding representation, which is based on deep learning, garners significant attention in practical applications~\cite{zhang2016collaborative,naumov2019deep,cheng2016wide}. Within the realm of recommendation systems, each specific model utilises its unique method for generating embeddings. As described in Section~\ref{sec:intro}, several approaches have been introduced, focussing on extracting multiple interests represented by item embeddings through the design of network structures. Furthermore, the influence of label diffusion into individual items within user interactive behaviour sequences diminishes the strength of user interests during the backpropagation process of the training model. While dynamic routing capsule networks can address this issue to some extent, integrating them specifically into recommendation models requires complex redesigning of the model architecture, rendering it a challenging and infrequently utilised approach. Based on DIEN~\cite{zhou2019deep}, the inclusion of an auxiliary task assumes a vital role in enhancing model performance. To avoid the need for redesigning the intricate architecture of the main recommendation model (DIN~\cite{zhou2018deep}), an auxiliary task is introduced to facilitate improved learning of item representations. Consequently, we propose a framework known as the User Interest-aware Capsule Network as an auxiliary task. Furthermore, the interest-based item embeddings derived from IaCN are shared with DIN~\cite{zhou2018deep}. Based on the formulation of $b_{ij}$, which employs the scaled dot product to measure the distance among interests and items, the item embedding vector is influenced by the user interest capsule vectors. Additionally, the auxiliary task incorporates the item embeddings expressed through user interest capsules into the main framework by sharing the item embedding vector. Within the main framework (as delineated in Figure~\ref{fig:recsys}). The item embeddings are illustrated as it consists of two distinct components, $\mathbb{E} = \mathbb{E}_{\text{original}}\oplus\mathbb{E}_{\text{auxiliary}}$, where the concatenation operator $\oplus$ is employed to combine the item embedding vectors, denoted as $\mathbb{E}_{\text{original}}$ and $\mathbb{E}_{\text{auxiliary}}$, respectively. The former corresponds to the item embedding vector in the primary target recommendation model task, while the latter is associated with the auxiliary task designed as IaCN. This framework enables the expansion of the original model's item embeddings by incorporating user interest capsules, while maintaining the original model architecture and preserving its inherent properties. As a result, this framework can be seamlessly integrated into a wide range of general recommendation models. The composition of the item embedding vector in the main recommendation model involves a combination of the original model's item embeddings and the auxiliary model's item embeddings. While the original item embeddings remain unaffected by the auxiliary task, the auxiliary item embedding is influenced by both tasks, with this influence being controlled through a hyperparameter. Consequently, the overall loss function for the entire model can be formulated as $\mathcal{L} = \mathcal{L}_{\text{DIN}}+\lambda\mathcal{L}_{\text{IaCN}}$, where $\mathcal{L}$ represents the overall loss of the complete model, $\mathcal{L}_{\text{DIN}}$ denotes the loss of the main model (deep interest network~\cite{zhou2018deep}, and $\mathcal{L}_{\text{IaCN}}$ corresponds to the loss of the user interest-aware capsule network. The hyperparameter $\lambda$ is used to fine-tune the balance between the loss and the auxiliary task loss. Given $\mathcal{L}_{\text{IaCN}}$, the label-aware attention layer necessitates positive samples for constructing the loss function. Consequently, negative samples corresponding to label items are masked to ensure that the loss function operates effectively during the training of the complete model. Throughout the model training process, the item embedding $\mathbb{E}$ receives two components of backpropagating gradients, one from the main model and the other from the auxiliary model. The original item embedding $\mathbb{E}_{\text{original}}$ solely receives the gradient from the main model $g^{e}_{\text{main}}$, while the auxiliary item embedding $\mathbb{E}_{\text{auxiliary}}$ receives gradients from both the auxiliary model (IaCN) $g^{e}_{\text{main}}$ and the main original model ($g^{e}_{\text{auxiliary\_main}}$) to facilitate its alignment ($\mathbb{E}_{\text{auxiliary}}$) with the main model. The update of the auxiliary model gradient is performed according to the illustrated formulation,
\begin{align}
    g^{e}_{\text{auxiliary}} = (1-\delta)g^{e}_{\text{auxiliary\_auxiliary}} + \delta g^{e}_{\text{auxiliary\_main}} \mid\delta\in [0,1]
\end{align}
The selection of the appropriate hyperparameter $\delta$ is detailed in the subsequent section.

\section{Experiments}
This section presents the experimental evaluation of the user interest-aware capsule network, which is introduced as an auxiliary task. The experiments are conducted using a real-world dataset from Amazon, and we detail the results and analysis in the following paragraphs.

\subsection{Datasets}
We employ the Amazon customer reviews, also known as product reviews dataset. The dataset utilised in this study comprises 192,403 consumer reviews obtained from Amazon electronics products, such as HDMI cables and Bluetooth speakers. A total of 63,001 products from 801 categories were covered, resulting in a dataset containing a total of 1,689,188 samples. The dataset encompasses essential product information, including details about each product, its corresponding category, and a list of viewed products. Another comprehensive collection of consumer reviews from Amazon books, encompassing 603,668 reviews in total are employed. These reviews pertain to 367,982 books, covering a wide range of 1,600 categories. In addition to the consumer reviews, the dataset provides essential product information, including details about the book's category and the list of viewed products.

\subsection{Implementation Details and Results}
We conduct experiments on the Amazon dataset encompassing electronics and books, and compare the performance of our model with other extensively employed baseline models within the industrial domain.\\
\textbf{Baselines:} We employ a set of baseline models for comparison, which are a) Wide and Deep~\cite{cheng2016wide}, it integrates a deep neural network (DNN) component as the `Deep' part and a linear model component as the `Wide' part. This integration aims to leverage the strengths of both models, facilitating a balance between model generalisation and memorisation. b) DIN~\cite{zhou2018deep} approach employs an attention mechanism to model the relationship between user interactive behaviour sequences and candidate items, enabling the determination of the contribution made by each item in the historical visiting sequences to the target item. And, c) DIEN~\cite{zhou2019deep} follows the architectures of DIN, where this approach incorporates gated recurrent units (GRUs) to effectively capture the evolutionary interests of users, while leveraging attention mechanisms.
\begin{table*}
   \caption{\label{tab:auc}Performance of individual models on the Amazon datasets. And, the analysis of hyperparameter $\delta$ for DIEN with the auxiliary network (IaCN) on Amazon books dataset.}
   \begin{tabular}{|c|c|c|c|c|}
    \hline 
       \multirow{2}{*}{\textbf{\diagbox{Model}{Dataset}}} &
       \multicolumn{2}{c|}{\textbf{Books}} & 
       \multicolumn{2}{c|}{\textbf{Electronics}} \\
      & AUC & SD & AUC & SD \\
      \hline
      DIN~\cite{zhou2018deep} & 0.7970 & 0.0010 & 0.7569 & 0.0009 \\
      \hline
      DIN~\cite{zhou2018deep}+IaCN & 0.8002 & 0.0009 & 0.7606 & 0.0013 \\
      \hline
      Wide and Deep~\cite{cheng2016wide} & 0.7860 & 0.0013 & 0.7461 & 0.0015 \\
      \hline
      Wide and Deep~\cite{cheng2016wide}+IaCN & 0.7928 & 0.0009 & 0.7502 & 0.0010 \\
      \hline
      DIEN~\cite{zhou2019deep} & 0.8534 & 0.0018 & 0.7706 & 0.0021 \\
      \hline
      DIEN~\cite{zhou2019deep}+IaCN & 0.8633 & 0.0019 & 0.7723 & 0.0002 \\
      \hline
   \end{tabular} 
   \begin{tabular}{|c|c|c|c|}
    \hline 
       \multicolumn{4}{|c|}{\textbf{DIEN~\cite{zhou2019deep}+IaCN}} \\
      $\delta$ & AUC & $\delta$ & AUC \\
      \hline
      0.1 & 0.862$_{\pm 0.0018}$ & 0.6 & 0.861$_{\pm 0.0014}$ \\
      \hline
      0.2 & 0.864$_{\pm 0.0019}$ & 0.7 & 0.862$_{\pm 0.0010}$ \\
      \hline
      0.3 & 0.864$_{\pm 0.0005}$ & 0.8 & 0.862$_{\pm 0.0018}$ \\
      \hline
      0.4 & 0.864$_{\pm 0.0013}$ & 0.9 & 0.862$_{\pm 0.0012}$ \\
      \hline
      0.5 & 0.861$_{\pm 0.0012}$ & 1.0 & 0.860$_{\pm 0.0022}$\\
      \hline
   \end{tabular}\\
\end{table*}
\begin{table}
    \centering
    \caption{Performance with different user interactive behaviour sequences length~$l$ on Amazon books}\label{tab:ablate}
    \begin{tabular}{c|c|c|c} \toprule
    Model & 10 & 20 & 50 \\
      \hline
       DIEN~\cite{zhou2019deep} & 0.854$_{\pm 0.0016}$ & 0.853$_{\pm 0.0018}$ & 0.850$_{\pm 0.0022}$ \\
       \hline
       DIEN~\cite{zhou2019deep}+IaCN & 0.857$_{\pm 0.0018}$ & 0.863$_{\pm 0.0019}$ & 0.861$_{\pm 0.0014}$ \\
       \hline
    \end{tabular}
\end{table}
To explore the performance enhancements, we introduce an auxiliary network (IaCN) that facilitates the extraction of multiple interests from the user's data. In this auxiliary network, we incorporate a layer of dynamic interest capsules, denoted as $K=[\log_2 |I^{c}_u|]$ to effectively extract the user's interests. The interest-based item embedding vector generated by the IaCN model is shared with the primary recommendation model, and its gradient is updated using a small ratio $\delta$=0.3. The overall loss is composed of both the main model loss and the auxiliary model loss, with the hyperparameter $\lambda$ set to 1 to appropriately weight the losses. The Adam optimizer is employed with a learning rate of 1$e$-4. Each experiment is replicated five times, and the standard deviation (SD) of the model's using the area under the curve (AUC) performance metric is computed to validate the effectiveness of our proposed framework. Table~\ref{tab:auc} presents the AUC of various baseline models. For the Amazon datasets, the maximum length of the list containing consumer-viewed electronic products is 10, while the maximum length for consumer-viewed books is 20. Notably, the framework incorporating the interest-aware capsule network as an auxiliary component consistently enhanced the performance of all model architectures, highlighting the advantages of this approach. Importantly, the main recommendation model retains its original structure while benefiting from the framework. Moreover, as the length of viewed items increases, the framework exhibits improved efficacy, particularly in the case of longer interactive behaviour sequences ($l$). To assess the performance of our proposed framework relative to the benchmark, we conducted experiments considering varying lengths of user-viewed sequences. The experimental results are presented in Table~\ref{tab:ablate}, indicating that DIEN with IaCN outperforms the original DIEN model in terms of AUC. Consequently, when dealing with long click sequences from users, the user interests capsule network demonstrates enhanced efficiency and accuracy in extracting diverse user interests, as verified through various experimental setups.

\section{Conclusion and Future Work}
In this work, we propose a comprehensive framework aimed at improving the performance of recommendation models through the utilisation of interest-based item embedding representations. To achieve this, a novel framework is introduced, which directly learns interest-based item representations by incorporating a user interest-aware capsule network. Furthermore, the user interest-aware capsule network is incorporated as an auxiliary component, facilitating the joint learning of both item-based item representations and interest-based item representations within the framework. Our framework can be readily applied to recommendation model architectures without the need for additional data or significant modifications to the model structure. Furthermore, the framework exhibits a notable impact on longer sequences of viewed items and a wider range of user interests. Through experimental evaluations on industrial datasets, it is evident that incorporating the user interest-aware capsule network as an auxiliary component outperforms other approaches in terms of click-through rate prediction tasks. Looking ahead, we intend to extend this work towards developing an explainable recommendation system that leverages the interest-aware capsule network to elucidate users' interests.


\bibliographystyle{ACM-Reference-Format}
\bibliography{sample-base}

\end{document}